\documentclass[journal=jacsat,manuscript=article]{achemso}

\graphicspath{{figs/}}
\usepackage{chemformula} 
\usepackage[T1]{fontenc} 
\usepackage{siunitx}
\usepackage{amsmath}
\usepackage{amsfonts}
\usepackage{graphicx}
\usepackage{mhchem}

\author{Antonia S. Kuhn}
\affiliation[MIT]
{Department of Materials Science and Engineering, Massachusetts Institute of Technology, Cambridge, Massachusetts 02139, United States}
\alsoaffiliation[ETH]{Department of Chemistry and Applied Biosciences, ETH Zurich, 8093 Zurich, Switzerland}
\alsoaffiliation[equal]{These authors contributed equally.}

\author{Jur\u{g}is Ru\v{z}a}
\affiliation[MIT]
{Department of Materials Science and Engineering, Massachusetts Institute of Technology, Cambridge, Massachusetts 02139, United States}
\alsoaffiliation[equal]{These authors contributed equally.}

\author{KyuJung Jun}
\affiliation[MIT]
{Department of Materials Science and Engineering, Massachusetts Institute of Technology, Cambridge, Massachusetts 02139, United States}

\author{Pablo Leon}
\affiliation[MIT]
{Department of Materials Science and Engineering, Massachusetts Institute of Technology, Cambridge, Massachusetts 02139, United States}

\author{Rafael G\'omez-Bombarelli}
\email{rafagb@mit.edu}
\affiliation[MIT]
{Department of Materials Science and Engineering, Massachusetts Institute of Technology, Cambridge, Massachusetts 02139, United States}

\title
  {Discovery of Polymer Electrolytes with Bayesian Optimization and High-Throughput Molecular Dynamics simulations}


\begin{document}
\maketitle

\begin{abstract}
    
Polymer electrolytes are critical for safe, high-energy-density solid-state batteries, yet discovering candidates that balance high ionic conductivity with high transference numbers remains a significant challenge. In this work, we develop a high-throughput screening platform that utilizes molecular dynamics simulations to navigate a chemical space of 1.7 million hypothetical polymer electrolyte candidates. Data from previous literature is used to warm-start batch Bayesian optimization for iteratively selecting new polymer electrolytes to evaluate. We iteratively identified, evaluated and analyzed 767 homopolymers as potential candidates. Our results reveal several candidates with transport properties exceeding the benchmark polyethylene oxide (PEO)/LiTFSI system. Crucially, our optimization campaigns for ionic conductivity and Li-diffusivity demonstrate that branched architectures and ketone functional groups significantly enhance ion-hopping mechanisms within the polymer matrix. We provide an in-depth mechanistic comparison of Li vs. Na cation transport and offer our open-source framework to accelerate the discovery of liquid, gel, and multi-cation electrolyte systems.
\end{abstract}


\section{Introduction}

Lithium-ion and lithium-metal batteries are essential for energy storage technologies in portable electronics, electric vehicles, and grid-scale storage. \cite{nicholaslim_electrolyte_2023} They dominate the current market due to their high energy density and long life cycle.\cite{duan_building_2020} However, conventional liquid electrolytes in Li-ion batteries present safety risks such as flammability when damaged, overheating during charging, and more. \cite{chen_review_2021, huang_questions_2021, wen_review_2012} Liquid electrolytes are susceptible to thermal runaway: a positive feedback loop of exothermic reactions that, once initiated, often leads to uncontrollable combustion and violent cell failure \cite{feng_thermal_2018}. Polymer electrolytes are a promising alternative due to their non-flammable nature, improved mechanical strength, and potential to suppress dendrite formation \cite{xiao_how_2019}. The underlying conduction mechanisms in polymer electrolytes have been investigated throughout experimental and computational work leading to the development of materials with increasing ionic conductivity and transference number. \cite{albertus_status_2018, mketkar_charging_2019, kim_status_2020, jun_lithium_2022, leon_mechanistic_2025} Despite this progress, achieving room-temperature ionic conductivity on par with liquid electrolytes remains a key challenge for polymer electrolytes.\cite{arya_polymer_2017, bocharova_perspectives_2020} 

Recent advances in machine learning (ML) and molecular simulations have accelerated the discovery and design of novel materials, including those used in energy storage systems\cite{gomez-bombarelli_design_2016, kirklin_high-throughput_2013, eames_ion_2014}. By enabling rapid screening of the vast chemical spaces, these computational approaches reduce the time and cost of discovering electrolyte candidates compared to traditional experimental methods. For example, using virtual screening in drug discovery has benefited the field in covering larger chemical spaces and cutting down costs. \cite{sabe_current_2021, sliwoski_computational_2014} In battery research, virtual screening approaches have allowed the discovery of new cathode, anode, and electrolyte materials for Li battery applications \cite{qu_identifying_2022, lun_cation-disordered_2021, liu_spinel_2015, jun_lithium_2022}. Molecular dynamics (MD) simulations have been particularly useful in electrolyte research by providing atomic-level insights into ion transport, polymer dynamics, and electrolyte-electrode interactions, bringing crucial information for optimizing performance and stability.\cite{brooks_atomistic_2018, genier_effect_2021, molinari_effect_2018, pesko_negative_2017,fong_ion_2021} MD simulations have shown to be capable of matching the kinetic properties of materials and thus, MD can be used as a proxy for experiments in a high-throughput discovery platform \cite{afzal_high-throughput_2021, ruza_benchmarking_2025}. 

Meanwhile, machine learning models trained on experimental and simulation data can predict material properties, identify key structure-property relationships, and guide the design of novel materials with enhanced properties.\cite{juan_accelerating_2021, tao_machine_2021, merchant_scaling_2023} ML based methods have enabled the discovery of electrode and electrolyte materials in Li based batteries and beyond. \cite{ahmad_machine_2018, moses_machine_2021, bradford_chemistry-informed_2023} Recent works have used MD simulations in tandem with generative models to create novel polymer candidates for Li-ion battery applications \cite{xie_accelerating_2022,khajeh_self-improvable_2024}. These works specifically concentrate on limited polymer space, and leave the door open for larger chemical exploration including more diverse chemical groups and larger variability in repeat unit sizes. 

In recent years,  Bayesian optimization (BO) has emerged as a powerful tool to accelerate the discovery of new materials. It is a method that iteratively explores a predefined search space balancing the exploitation-exploration trade-off with the purpose of finding the optimal set of parameters or candidate in the said search space. It has been used in various fields including the discovery of alloys with higher mechanical strength, identifying novel electro-catalyst candidates for \ce{CO2} reduction, and the oxygen evolution reaction. \cite{ghorbani_active_2024, tran_active_2018, jjenewein_navigating_2024} These methods have also been applied in the discovery of aqueous and non-aqueous Li-ion battery electrolytes. \cite{dave_autonomous_2020, dave_autonomous_2022}. While traditionally used in a sequential manner, batch BO methods have been developed to sample multiple candidates at the same time in cases where parallel processing capabilities are available\cite{wu_parallel_2016, gonzalez_batch_2016}. Furthermore, it has been shown that warm-starting the BO campaign with either literature data or transferring knowledge from similar domains accelerates the discovery of best candidates \cite{jhickman_equipping_2023, ruza_autonomous_2025}.

\begin{figure}[h!]
    \centering
    \includegraphics[width=0.90\textwidth]{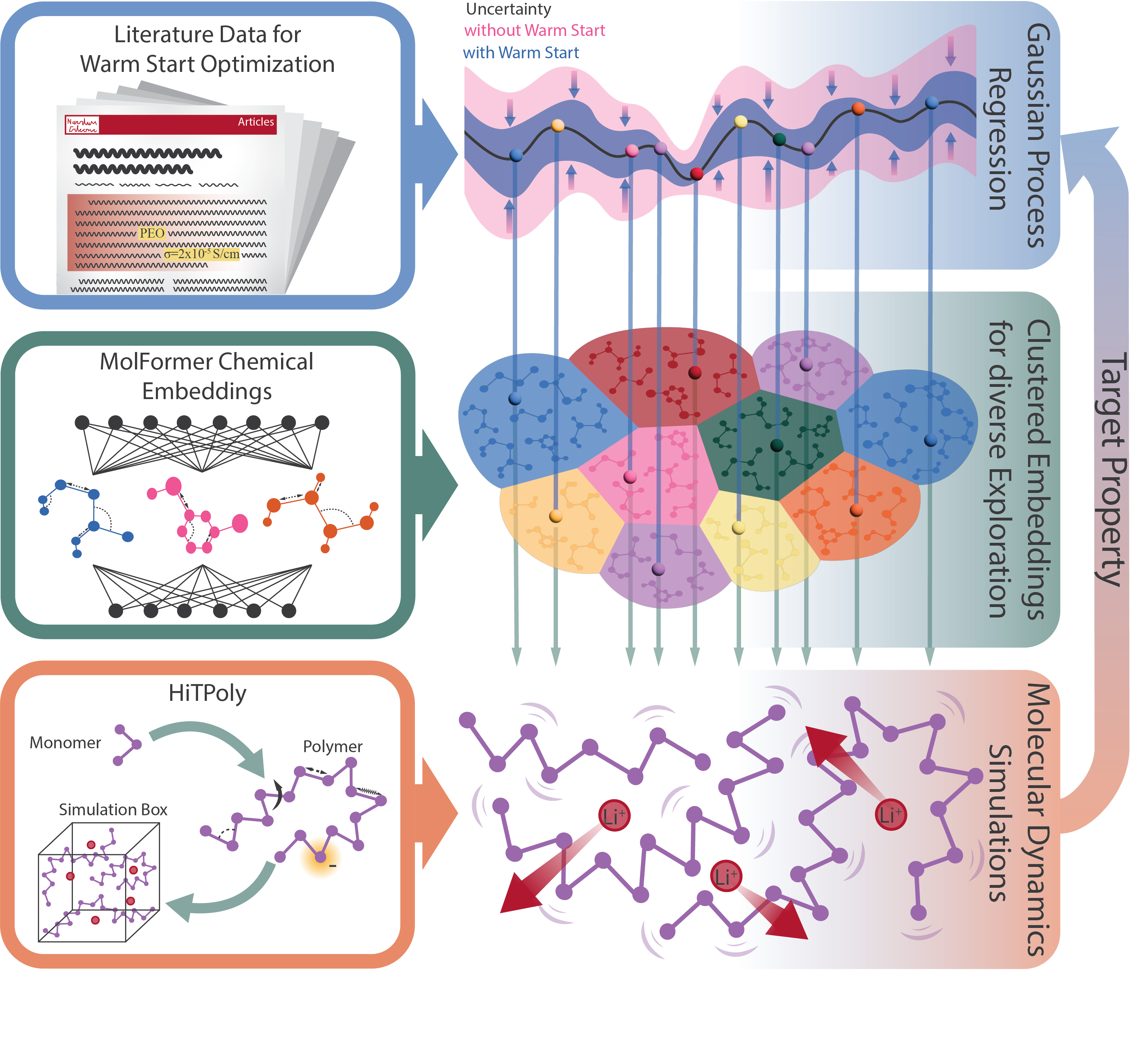}
    \caption{Workflow of our active learning high-throughput pipeline. Experimental values from literature are used for a warm start of the Gaussian Process Regression (GPR) model. The polymers are embedding with the MolFormer model and this space is clustered to enforce larger exploration of the chemical space. The HiTPoly workflow is employed to simulated the sampled polymers and obtain ionic conductivity to retrain the GPR model and sample new candidates for multiple batches.}
    \label{fig1}
\end{figure}

In this work, a Bayesian Optimization framework is proposed to find the best polymer electrolyte candidate (Figure \ref{fig1}). The Open Macromolecule Genome \cite{kim_open_2023} and Small Molecules into Polymers \cite{ohno_smipoly_2023} datasets of synthetically accessible polymers are used for screening. Polymer candidate chemical structure is captured with MolFormer embeddings, a large-scale chemical language model. \cite{ross_large-scale_2022} Gaussian process regression is then used as the surrogate model for predicting ionic conductivity, transference number, ion clustering and Li diffusivity. Previously collected experimental literature data is employed in a warm start for the BO campaign. \cite{bradford_chemistry-informed_2023} Each selected polymer candidate is investigated with molecular dynamics simulations to obtain the ionic conductivity, transference number, solvation environments and ion and polymer diffusivity. This pipeline is used to screen the chemical space of 1.7 million candidates and find polymer electrolytes with ionic conductivity higher than that of PEO. Finally, 10 polymers with high ionic conductivity and diverse backbone chemistry are selected for transport mechanism decomposition investigation.

\section{Methods}


A dataset of 1.66 million homopolymer candidates was constructed for virtual screening from theoretical databases that utilized chemical heuristics to propose polymers that are likely synthesisable. The majority of the dataset consists of the Open Macromolecular Genome\cite{kim_open_2023}, with a small addition from the Small Molecules into Polymers (SMiPoly)\cite{ohno_smipoly_2023} dataset. The largest database of synthesized polymers, PoLyInfo, consists of around 19,000 homopolymers, which is significantly smaller than our screening space. The resulting homopolymer search space consisted of polyolefins, polyesters, polyethers, polycarbonates, polyamides, organic sulfides, and others giving a wide range of materials to study for lithium-ion (and sodium-ion) conducting electrolytes. 

Bayesian Optimization was used to navigate the large search space in an efficient manner. We increased optimization performance by supplying the surrogate model with experimental data for a warm start to the campaign. Ionic conductivity of 106 homopolymers with the LiTFSI salt was extracted for salt concentrations in the range 0.2––2~M and temperatures from 70--90~$^{\circ}$C from the experimental database compiled by Bradford et al.\cite{bradford_chemistry-informed_2023} Joining experimental and computational (simulated at shifted temperature \cite{ruza_benchmarking_2025}) data the surrogate model performance is not affected (SI section 3). Other recent work has shown increased machine learning model performance when mixing computational and experimental data in polymer applications. \cite{toland_accelerated_2023}

The polymers were encoded into a SMILES representation with multiple CRUs such that the representation of each polymer has around 200 atoms, where the ends of the polymer chain are capped with a methyl group. The chemical large language model, MolFormer\cite{ross_large-scale_2022}, was applied to generate embeddings, resulting in a 768-long continuous vector for each polymer. Embeddings serve as dense numerical descriptors that capture latent physicochemical properties and structural topology, translating the discrete molecular representation into a continuous feature space \cite{duvenaud_convolutional_2015}. While language models such as polyBERT \cite{kuenneth_polybert_2023} exist for polymer molecule embeddings, we chose MolFormer because it was trained on a larger corpus of chemical data. The Gaussian Process (GP) surrogate model used in optimization was trained on 106 experimental data points, which is much smaller than the feature space. GPs are known to perform poorly on high dimensional data through distance behavior in high dimensions \cite{aggarwal_surprising_2001}. Therefore, dimensionality reduction was performed with principal component analysis (PCA) \cite{abdi_principal_2010} on the MolFormer features scaled with a standard scalar to reduce the embeddings to 50 features which explain $\sim$80~\% of the variance (see SI section 1).  Gaussian process regression with a Radial-Basis-Function kernel of lengthscale 1 and an output scale of 0.5 was trained on the PCA features and the experimentally-measured ionic conductivities. For a comparison of different chemical representations and regression models, refer to SI section 1. During the BO campaign we used the expected improvement acquisition function\cite{snoek_practical_2012} to suggest the next samples. 

The search space was clustered for each batch with k-means++ based on its Manhattan distance into 20 clusters. During each optimization batch one polymer is sampled from each cluster to increase chemical diversity (SI Figure S4). \cite{arthur_k-means_2006} The polymers with the highest expected improvement from each cluster were sampled for each batch. The simulated cluster Nernst-Einstein ionic conductivities were added to the training data, and the prior model belief was updated \cite{arthur_france-lanord_correlations_nodate}. 

In an effort to account for other factors that impact the performance of batteries, such as, transference number, fraction of free charge carriers or ratio of lithium diffusivity to CRU diffusivity, more batches were run to screen for these properties. High transference number is important as it allows more charge being carried by Li(Na)-ions and prevents the formation of charge gradients in the battery. \cite{choo_diffusion_2020} The fraction of free charge carriers works as a proxy for ion solvation, which is important to be able to pack a higher amount of ions in the electrolyte \cite{ruza_benchmarking_2025}. Finally, a high ratio of lithium diffusivity to CRU diffusivity can indicate that lithium transport is decoupled from polymer segmental motion \cite{ruza_benchmarking_2025}. We choose these properties because they allowed us to explore alternative avenues for increasing polymer electrolyte performance. Due to a lack of experimental data for these properties, only simulation data was used for training the GP surrogate model. All factors of the representation, model and simulation stayed consistent. We screened for the lithium transference times the fraction of free charge carriers for five more batches and obtain 88 new simulations. Five batches were run that optimized for lithium transference times the ionic conductivity, and another 114 new polymers were screened. Finally, we screened for the ratio of lithium diffusivity to CRU diffusivity times the ionic conductivity for another five batches (99 new polymers screened). In total, we simulated 767 polymer electrolyte systems. 

All simulation files were built and analysis was performed with HiTPoly \cite{ruza_benchmarking_2025}. We obtain OPLS-AA force field parameters from the LigParGen.\cite{jorgensen_potential_2005, dodda_114cm1a-lbcc_2017, dodda_ligpargen_2017} The ion charges were scaled with a factor of 0.75 in order to reproduce experimental structural and transport properties of alkaline salts in organic electrolyte systems in alignment with previous studies that suggest a factor of 0.7--0.8 to account for the lack of polarizability in classical MD.\cite{molinari_transport_2019, mogurampelly_structure_2017, costa_polymer-ionic_2015} Simulations were run at 393~K instead of the same temperature of the experimental pre-trained dataset range of 343-363~K. Previous work by Ruza et al. found that shifting the simulation temperature by 30-50~K (the difference in the glass transition temperature between experiments and computations) brings a higher match to experiments for class I force fields. \cite{ruza_benchmarking_2025} 
This aligns with similar results by Fang et al. \cite{fang_molecular_2023} where a temperature shifting was proposed by the difference between the experimental and computationally calculated glass transition temperature. Furthermore, the 50~K offset determined by Ruza et al. aligns with an offset reported by Atif Faiz Afzal et. al.\cite{afzal_high-throughput_2021}. Experimental and computationally calculated glass transition temperatures were compared in a wide variety of polymers and an average offset of 50~K was found. The ionic conductivity was computed with the cluster Nernst-Einstein approach (cNE) \cite{arthur_france-lanord_correlations_nodate} which approximately accounts for the clustering of ionic species in polymer electrolytes in a static manner. The work of Ruza et al., established that for high ionic conductivity polymers ($>10^{-6}$ S/cm) cluster Nernst-Einstein ionic conductivity and transference numbers converge within 100 ns of simulation time.\cite{ruza_benchmarking_2025} The concentration of the salt was always set to 1~M. For further details on the simulations and property calculations (SI section 4).


\section{Results and Discussion}

\begin{figure}
    \centering
    \includegraphics[width=0.9\linewidth]{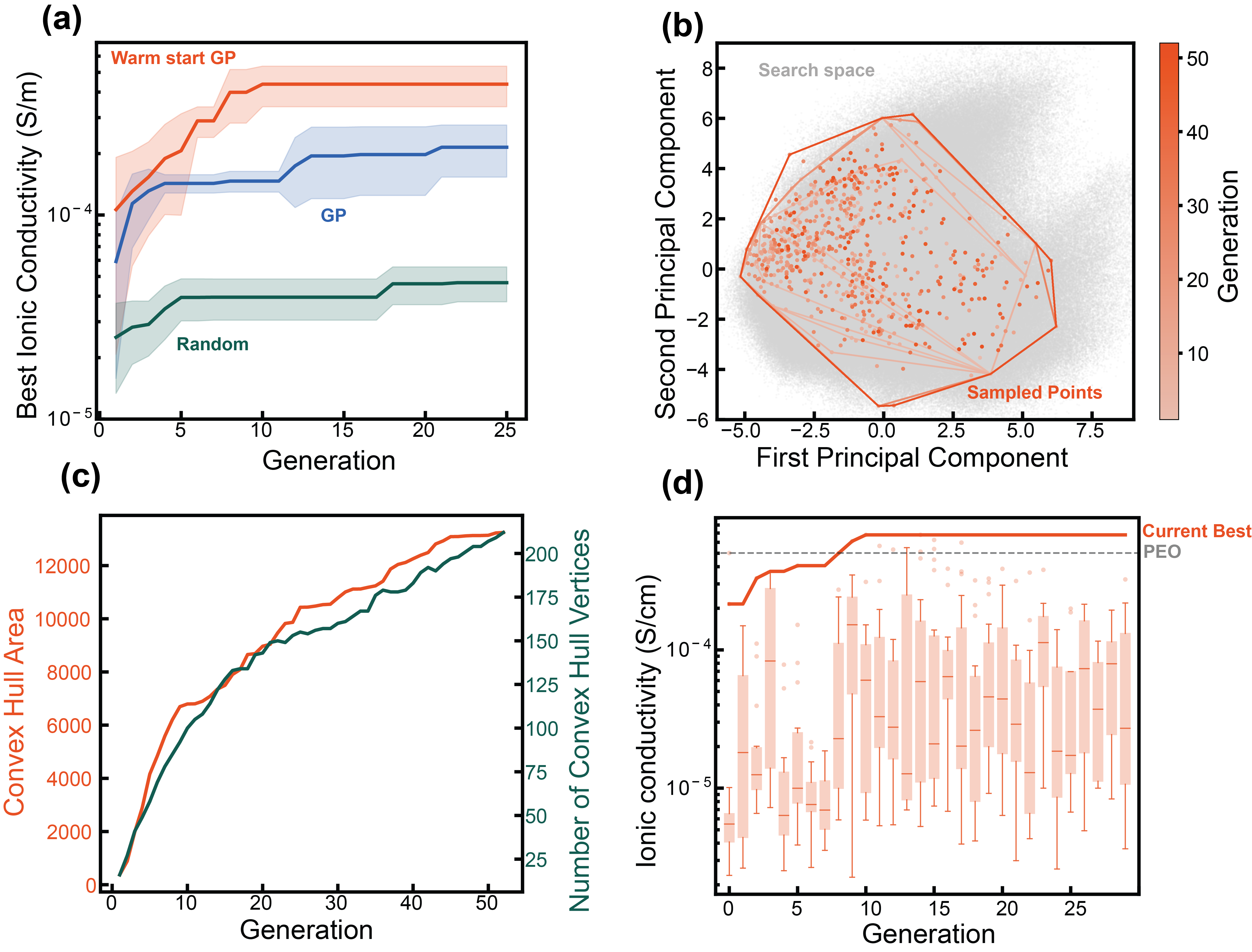}
    \caption{(a) Performance comparison of our warm start GP model against a no warm-up GP and random search for 25 generations. (b) The first two principal components of the total search space in gray with the evolution of the convex hull by batch of training. (c) The evolution of convex hull area and number of convex hull vertices as a function of generation run for the first 5 principal component dimensions. (d) The evolution of the best ionic conductivity for each batch with boxes representing the mean and variance of the ionic conductivity for the whole batch.}
    \label{fig:panel2}
\end{figure}

To evaluate the proposed workflow, we retrospectively compared the warm-start GP against both a cold-start GP and random search. To facilitate this evaluation, we trained an oracle model on the complete simulation dataset to serve as the ground truth for ionic conductivity (details in SI Section 1). Figure \ref{fig:panel2} (a) shows the comparison of the three methods. The warm start GP significantly outperformed the GP and the random search in the 25 test batches (with 20 suggestions at each generation, emulating the optimization campaign). This shows that utilizing experimental data to warm-start computational screening campaigns gives an advantage over methods without prior knowledge in finding samples with higher ionic conductivity in fewer iterations.

Figures \ref{fig:panel2} (b) and (c) show the benefit of using MolFormer molecular embeddings that encode molecular structure information for chemical space exploration. The gray points in Figure \ref{fig:panel2} (b) show the first two dimensions of PCA of the MolFormer embeddings (of the 50 features that were used for the training the GP), and the orange points show the evolution of the candidates with increasing batches of the optimization campaign. The orange lines show the convex hull of the first 2 dimensions of search space embeddings. Figure \ref{fig:panel2} (c) shows the evolution of the convex hull and the number of convex hull vertices during the optimization run of the first five PCA dimensions (2,3,7,10 dimension figures available SI section 6, the trends remain the same). It can be seen that, with increasing generation amount, the model expands the convex hull area and continues to explore a wider variety of different polymer chemistries, balancing exploration and exploitation.

Figure \ref{fig:panel2} (d) shows the evolution of cNE ionic conductivity throughout the campaign while maximizing ionic conductivity. Each histogram shows the mean and standard deviation of the cNE ionic conductivity for the 20 simulated polymers. For the first 10 batches, there was an increase in maximum cNE ionic conductivity for each batch. In batch 8 a polymer with cNE ionic conductivity as high as PEO was found, and in batch 10 the polymer with the highest cNE ionic conductivity of the campaign was found. During the next 20 batches, more polymers with cNE ionic conductivity similar to that of PEO and the current highest cNE ionic conductivity were found. The absence of significant performance gains in later iterations implies that the high-performing regions of the chemical space have been effectively saturated.

\begin{figure}
    \centering
    \includegraphics[width=0.9\linewidth]{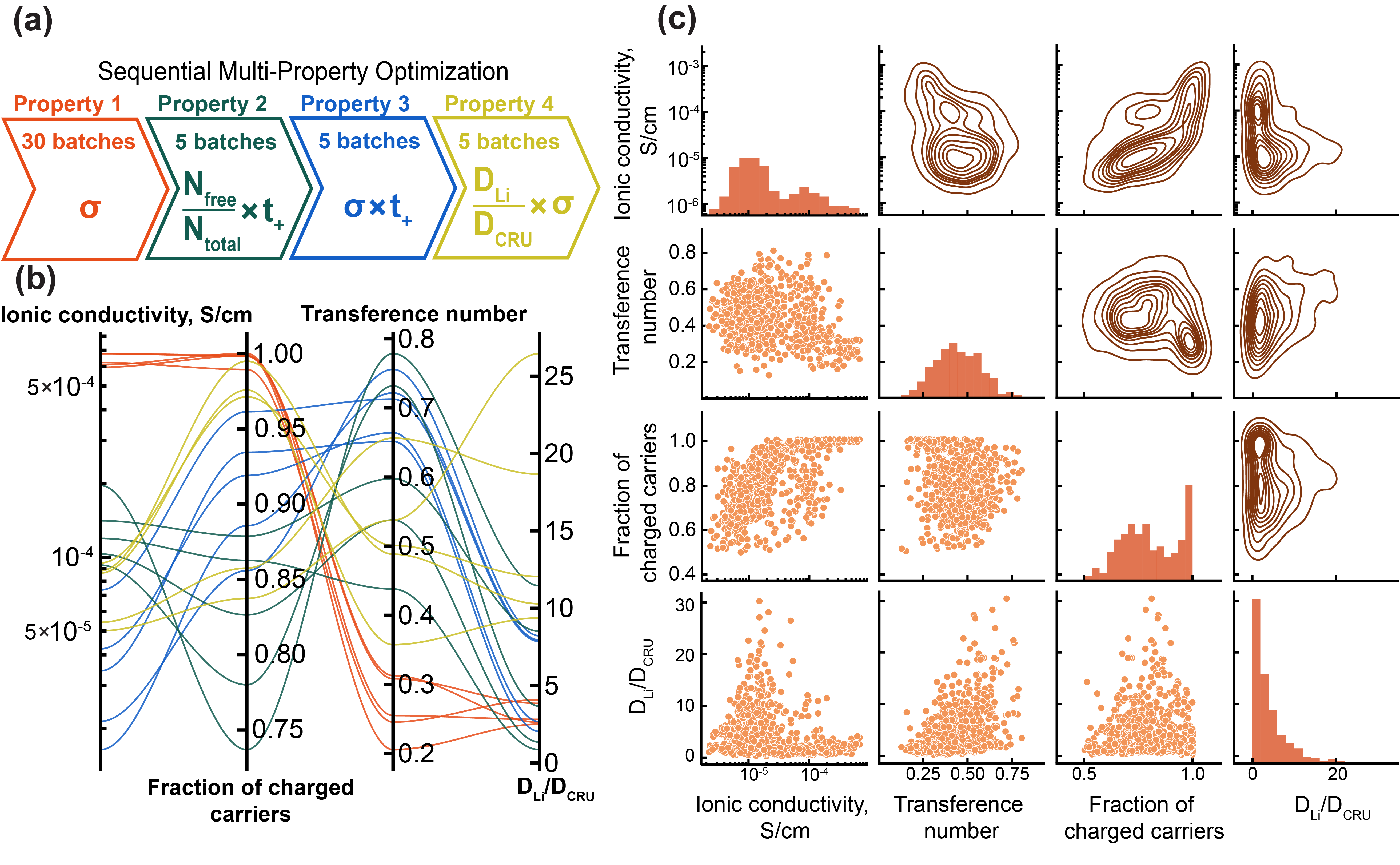}
    \caption{(a) Schematic of the different optimization targets during this campaign. (b) Parallel coordinates of the five polymers with the highest performance for each optimization property decomposed by the four main components of the properties. (c) All pairwise relationships between the four properties of interest. Diagonal plots show the distribution of each property, lower triangle shows the pairwise scatter plots and the upper triangle shows the bivariate distributions using kernel density estimation.}
    \label{fig:panel3}
\end{figure}

After 30 batches of optimization for ionic conductivity (property 1), other properties were selected to be optimized. Figure \ref{fig:panel3} (a) shows the properties and the number of batches for which the property is optimized. The fraction of charged carriers times the transference number was chosen as the second property (property 2). The fraction of charged carriers is defined by the number of ions in charged clusters divided by the total number of ions in the system and can be seen as a proxy of the solvation capability of Li-ions and the transference number is an important property for Li-ion electrolytes. The work of Ruza et al. \cite{ruza_benchmarking_2025} showed that there is an inverse trend between the transference number and the fraction of charged carriers, thus optimizing for the increase of both properties could lead to polymers with both high transference number and high cation solvation. The third property to optimize was the cNE ionic conductivity multiplied by the transference number, as finding a polymer with both a high ionic conductivity and high transference number would be beneficial (property 3). Finally, the fraction of lithium-ion self-diffusivity and constitutional repeat unit (CRU) self-diffusivity multiplied by the cNE ionic conductivity (property 4) was optimized. Achieving high ratio between the Li-ion and CRU diffusivity could imply that the the mechanism of ion conduction is uncoupled from that of the polymer chain dynamics. 

Figure \ref{fig:panel3} (b) shows the trade-off of the four properties that make up the four optimization targets for the top 5 candidates in each optimization run (structures of these polymers available in SI section 6). It can be clearly seen that when optimization is performed for property 1, a side effect is achieving a high fraction of charged carriers (highly dissociated Li ions), yet the transference number and the diffusivity ratio are low. 
When optimizing for property 2 the model found multiple polymers with high transference number, while sacrificing some of the fraction of charge carriers. Overall, there is a large spread in both properties for this optimization target, with transference number being as low as 0.44 for one polymer candidate and as high as 0.78 for another polymer candidate. Optimizing for property 3, the model found candidates with a consistently high transference number while suffering from lower ionic conductivity compared to the other polymers. Interestingly, these samples exhibit a high fraction of charge carriers, which is unusual when there is a high transference number. \cite{ruza_benchmarking_2025} Finally, by optimizing for property 4 we discovered some polymers with very high diffusivity ratio while sacrificing an order of magnitude of cNE ionic conductivity compared to the polymers with the highest cNE ionic conductivity.

To evaluate the overall trends between the four properties in the optimization targets that encompass all batches, a correlation matrix is shown in Figure \ref{fig:panel3} (c). The lower triangle of the matrix shows scatter plots between each property, the diagonal displays the distribution of each property, and the top triangle shows the probability density distribution of each property pair. The distribution of ionic conductivities shows a bimodal distribution with a higher density at the low conductivity range. One possible cause for this is that the ionic conductivity in the range $<10^{-5}$ S/cm is overrepresented as MD simulation of 100 ns is not long enough to obtain enough sampling to compute well-converged ionic conductivities of slow conductors with ionic conductivities below $<10^{-6}$ S/cm. The distribution of fraction of charge carriers also follow an interesting trend, that the majority of candidates follow a mostly normal distribution, with a large peak at the maximum fraction of charge carriers of 1. As seen in Figure \ref{fig:panel3} (b), when optimizing for high ionic conductivity, the samples often also follow a very high fraction of free charged carriers. Having a high fraction of free charged carriers implies high solubility of the ionic species. The more ions are participating in the conduction mechanism, the higher the ionic conductivity. Since the majority of the optimization campaign was focused on finding the candidate with the highest ionic conductivity, it follows that many of the samples would have a high fraction of charged carriers. 

\begin{figure}
    \centering
    \includegraphics[width=0.9\linewidth]{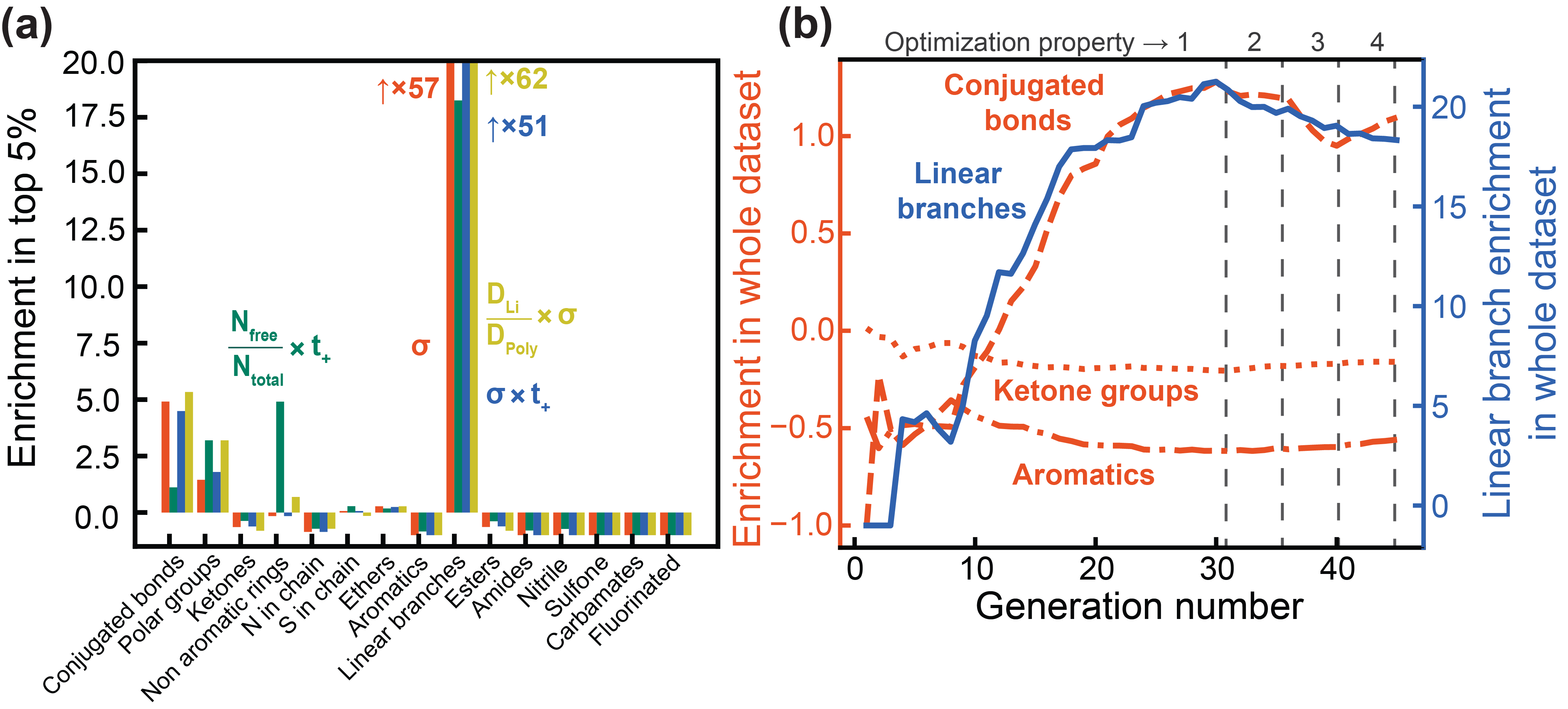}
    \caption{(a) Enrichment of different types of chemical groups for the various optimization targets in the top 5\% best samples for each target compared to the prevalence of those chemical groups in all of the tested samples. (b) Enrichment evolution for linear branches (blue, right axis), conjugated groups (dashed line), ketone groups (dotted line), aromatic groups (dotted-dashed line) across the 45 batches compared to prevalence of those groups in the total dataset.}
    \label{fig:panel4}
\end{figure}

To translate the optimization results into actionable chemical insights, we employed a fragmentation scheme that decomposed each polymer from the theoretical database, allowing us to identify the key chemical motifs correlated with high performance. For example, polymers that had double bonds in their structure were marked as having a conjugated bond, polymers that had a R-C(=O)-R' group with R, and R' being any substituent were marked as having ketone groups. Polymers that contain linear branches were defined as those that did not include any 5 or 6 member rings in their structure and had at least one branch/bottlebrush-like structure of at least 5 heavy (C, O, N, S, etc.) atoms. These linear branch polymers are similar to bottlebrush polymers that have shown to have good mechanical properties and high room temperature ionic conductivity in gel form. \cite{jia_multifunctional_2021} To gain insight into whether the model was able to uncover underlying chemical information from the dataset the enrichment was tested in the top 5\% of candidates and across the entire dataset. Enrichment in the top 5\% ($E_{top5}$) was defined as:

\begin{equation}
    E_{top5} = \frac{f_{top5} - f_{all}}{f_{all}},
\end{equation}

where $f_{top5}$ is the fraction of candidates with a certain chemical group in the top 5\% of the sampled candidates, and $f_{all}$ is the prevalence of these chemical groups in the whole dataset. The enrichment ($E_{total}$) in the theoretical polymer database was defined as:

\begin{equation}
    E_{total} = \frac{f_{campaign} - f_{all}}{f_{all}},
\end{equation}

where $f_{campaign}$ is the prevalence of that chemical group in the optimization campaign. Figure \ref{fig:panel4} (a) shows the enrichment in the top 5\% in a wide variety of chemical groups. The overall prevalence of these chemical groups in the entire dataset is shown in SI Figure 8. The main takeaway is that polymers with linear branches were observed 57 times more in the top 5\% of candidates optimized with ionic conductivity compared to the rest of the search space, 51 times more when optimized for ionic conductivity times transference number, and 62 times more prevalent when optimizing for diffusivity ratio times ionic conductivity. This is heavily emphasized in the fact that of the 30 polymers with the highest screened ionic conductivity, 27 have a structure with linear branches (full list of screened polymers with all the properties in the SI). Across all four optimization targets an enrichment of conjugated bonds and polar groups was noticed, pointing towards the beneficial nature of these chemical groups to high performance in the optimization of the 4 properties. Finally, the enrichment in the top 5\% of non-aromatic rings for the fraction of free charged carrier time transference number is surprising. We hypothesize that these polymer repeats units that have multiple polar groups attached to the rings (examples in the SI section 6), can arrest the movement of the anion and thus increase the transference number. 

Figure \ref{fig:panel4} (b) shows the search space enrichment as a function of the optimization batch number. Firstly, the prevalence of ketone groups stayed mostly constant throughout the optimization campaign, which more than anything speaks to the fact that ketones are prevalent in most of the search space (>90\% of all polymers in the dataset have a ketone group). Secondly, the model starts immediately with a negative enrichment of aromatic groups in the optimization campaign towards ionic conductivity. This shows the power of having instructed the model with warm start literature data, as this gives prior knowledge that aromatic groups are unlikely to lead to an increase in ionic conductivity for polymer electrolytes. Finally, for both linear branches and conjugated bonds there is a similar trend, albeit with an order of magnitude difference in their intensity, where in the first couple of generations they were sampled less prevalent than they are in the whole dataset, but as the model discovered their benefits to the properties these molecules were sampled to a higher extent.

\begin{figure}
    \centering
    \includegraphics[width=0.9\linewidth]{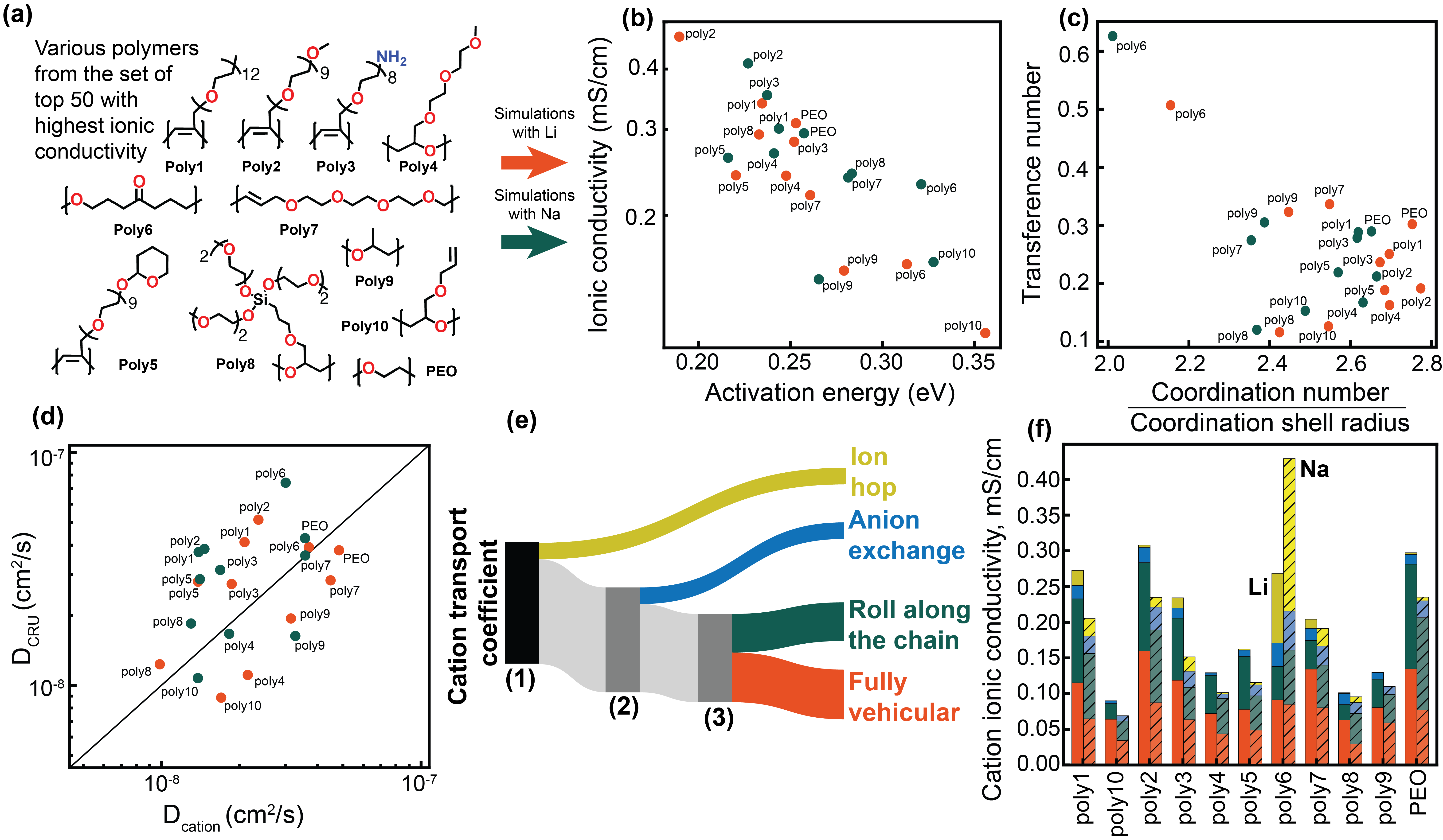}
    \caption{(a) 11 polymers with varying structures selected from the top 50 polymers with the highest ionic conductivity. (b)  Ionic conductivity at 393 K against the activation energy of ionic conductivity for Li (orange points) and Na (green points). (c) Transference number against the ratio of coordination number and coordination shell radius both at 393 K. (d) Simulated NE Li diffusivity against CRU diffusivity both at 393 K. (e) Sankey diagram of the decomposition of the various transport mechanisms. (1) Polymer chains entering or leaving solvation shell (top yes, bottom no). (2) Anions entering or leaving the solvation shell (top yes, bottom no). (3) Change in coordinating atoms from the majority solvating polymer chain (top change in coordinating atoms, bottom no change). (f) The contributions of each of the transport mechanisms to the overall cation ionic conductivity for each of the polymers at 393 K. }
    \label{fig:panel5}
\end{figure}

After the whole optimization campaign was completed, 10 (11 with PEO) polymers with high ionic conductivity and distinct chemical structures were selected for more in-depth testing. Figure \ref{fig:panel5} (a) shows the structures of the 10 (+ PEO) polymers selected for additional testing. Simulations were performed with both Li and Na as the cations and TFSI as the anion at various temperatures to calculate activation energies of these polymers. The relationship between ionic conductivity and ionic conductivity activation energy can be observed in Figure \ref{fig:panel5} (b). A negative linear trend between the two variables can be seen for both Li and Na polymer simulations. Most of the polymers have very similar results for both Li and Na variations. Figure \ref{fig:panel5} (c) demonstrates the relationship between the cation transference number and the ratio between the coordination number and coordination shell radius (decomposed figure for the coordination number and coordination shell radius in SI Figure S11). The coordination shell radius was obtained by calculating the mean bond length of the convex hull of the solvation environment. The majority of the polymers are clustered together in a similar range of transference numbers and ratios between coordination number and coordination shell radius, except poly6, which is a strong exception compared to PEO and the other polymers. Figure \ref{fig:panel5} (d) shows the relationship between cation and CRU diffusivity. Of the selected 10 polymers about half align above the diffusivity parity line, while the other half below. Overall, for all of these polymers the ratio between diffusivities is close to 1, which implies a correlated cation motion with the motion of the polymer chain.

To obtain deeper insights into what makes some polymers have a higher ionic conductivity or transference number, an in-depth transport mechanism analysis was performed following the algorithm developed by Jun et al.\cite{jun_universal_2026} The general algorithm works by slicing the MD trajectory by a given frequency (100 ps in this work) \cite{leon_mechanistic_2025}, for each lithium (sodium) assigning each window and its corresponding displacements to one of the four predefined types of events based on how the solvation environment of lithium changes during each time window. This information is aggregated to compute the exact contribution of each event towards Li transport property. Figure \ref{fig:panel5} (e) shows the algorithm to assign each time window of each lithium to one of the transport mechanisms. The first split in Figure \ref{fig:panel5} (e) (1) evaluates if there are polymer chains that enter or leave the cation solvation shell in some $\Delta t$. The yellow line on the top marks an ion hop, that is, in some $\Delta t$ there is an interchain hop of the cation. The bottom split is when there is no change in polymer chain index (or residue) in the solvation environment. The (2) split denotes if there are anions (TFSI) leaving or entering the solvation environment, with the blue part denoting an anion exchange event, when one anion leaves and another enters the solvation shell. The final split denotes if there is a change of solvating atom indices in the solvation shell (O, S, N) from the majority polymer. The majority polymer is defined as the residue providing the largest number of coordinating atoms during its residence time. If the majority polymer changes (top green part), it is defined as a roll along the chain (or intrachain hop). No changes in the solvation environment (bottom orange part) are classified as a fully vehicular transport mechanism. Figure \ref{fig:panel5} (f) demonstrates the contributions of each mechanism to the overall cation ionic conductivity. The sum of the contribution of each mechanism adds exactly to the total cation Nernst-Einstein conductivity. Similarly to how in Figure \ref{fig:panel5} (c) the majority of the polymers are clustered together, in Figure \ref{fig:panel5} (f) most of them have very similar contributions from the various mechanisms. An exception is poly6, that has a significantly higher ion hopping component than the rest of the polymers. The work of Jun et al. \cite{jun_universal_2026} has showed in LiTFSI + PEO that when normalizing the decomposed ionic conductivity of lithium for each mechanism by its corresponding probability of occurrence, interchain hop-related events are the most effective mode of Li transport in PEO. This implies that it is important to target polymer electrolytes where there is a significant contribution of ion hopping to the overall cation transport mechanism as in the case of poly6. Furthermore, poly6 is similar in structure to Polycaprolactone and Poly(pentyl malonate) (PPM) that have both shown high transference numbers and similar properties to poly6. The downside of these polymers with ketone groups is the lower fraction of free charge carriers (Figure S12 in the SI) which implies lower than PEO cation solvation.

\section{Conclusion}

In this work, we used a high-throughout molecular dynamics simulation pipeline (HiTPoly) that has been integrated into a Bayesian optimization framework with chemistry-informed molecular embeddings for the discovery of novel polymer electrolytes. We showed that using a warm-start Gaussian Process (GP) outperforms an out-of-the box GP even when trained on experimental data, and with increasing batches the model keeps sampling more diverse chemical space. Importantly, during the optimization campaign we found multiple polymer electrolytes with ionic conductivity higher than that of PEO. We pursued four different metrics that affect the electrolyte performance, ionic conductivity, transference number, fraction of charged carriers as well as the ratio of lithium-ion to CRU-unit diffusion. This approach allowed us to investigate polymers with different desirable properties, paving the way for a deeper understanding of polymer electrolytes as a whole. 

Examination of the impact of chemical groups on the screened polymer candidates revealed that the model heavily over-sampled linear branches (side-chains of >5 heavy atoms from the polymer backbone) compared to the rest of the chemical groups. Out of the top 30 polymer electrolytes with the highest ionic conductivity 27 had a linear branch in the structure. These linear branches are similar to that of bottle-brush polymers, except smaller in size. Making polymers as such could be beneficial for next-generation polymer electrolytes.
Deeper analysis was performed on 10 chemically distinct candidates from the top 50 highest ionic conductors. The analysis revealed that the polymers with linear branches had similar solvation properties to PEO as well as comparable ion conduction mechanisms. Ketone based chemical groups in the polymer backbone contributed significantly to having a larger ion hopping contribution to the overall cation Nernst-Einstein conductivity and points to the importance of promoting interchain hopping events to improve cation conductivity, which is a sought after property for polymer electrolytes.

Overall, more innovations are needed in the polymer electrolyte space to find the optimal candidate for li(na)-ion(metal) batteries. The hypothetical polymer dataset screened in this work was not exhaustive but it successfully identified candidates with performance exceeding that of PEO. This finding suggests that a promising path forward is the optimization of novel architectures, such as branched polymers with ketone groups, which may enhance cation conductivity by facilitating an ion-hopping mechanism. It has to be noted that incremental increases in conductivity and transference number for homopolymers are difficult and we need to combine these guiding principles with novel architectures like copolymers or single-ion conducting polymers for the next generation of electrolytes for safer and higher performance batteries.

\section{Acknowledgment}

This work was supported by the Toyota Research Institute. The authors acknowledge partial support from the Energy Storage Research Alliance "ESRA" (DE-AC02-06CH11357), an Energy Innovation Hub funded by the U.S. Department of Energy, Office of Science, Basic Energy Sciences. We acknowledge the MIT Lincoln Laboratory Supercloud clusters as well as computational resources of the MIT Office of Research Computing and Data. The authors acknowledge Professor Yang Shao-Horn and Professor Jeremiah Johnson for discussions during the project.

\section{Code and Data availability}

The code used in this workflow is available on github \url{https://github.com/learningmatter-mit/HiTPoly}. The full list of all polymers tested and their molecular dynamics analyzed properties are available at 10.5281/zenodo.18674321.

\bibliography{references}

\end{document}